# In 'Private, Secure & Conversational' FinBots We Trust

Magdalene Ng[†], Kovila P.L. Coopamootoo[†], Tasos Spiliotopoulos[‡], Dave Horsfall[†], Mhairi Aitken[‡], Ehsan Toreini[†], Karen Elliott[‡], Aad van Moorsel[†]

[†]School of Computing, Newcastle University; [‡]Newcastle University Business School

## ABSTRACT

In the past decade, the financial industry has experienced a technology revolution. While we witness a rapid introduction of conversational bots for financial services, there is a lack of understanding of conversational user interfaces (CUI) features in this domain. The finance industry also deals with highly sensitive information and monetary transactions, presenting a challenge for developers and financial providers. Through a study on how to design text-based conversational financial interfaces with N=410 participants, we outline user requirements of trustworthy CUI design for financial bots. We posit that, in the context of Finance, bot privacy and security assurances outweigh conversational capability and postulate implications of these findings. This work acts as a resource on how to design trustworthy FinBots and demonstrates how automated financial advisors can be transformed into trusted everyday devices, capable of supporting users' daily financial activities.

## INTRODUCTION

CUIs in different contexts are widely and substantially studied, such as in the area of health management, home environments and even conflict research [3,9,10]. These features typically gain positive responses and are preferred to technical and mechanical bots in these aforementioned domains. Social and conversational features here mean that the bot uses polite language, adheres to conversational turn-taking, makes small talk, greets users with their preferred name and even provides individual feedback [15]. Increasingly, financial firms and platforms are commercially introducing the creation of socially-apt financial bots, or FinBots[1]. FinBots are a specific type of automated software application created to automate certain tasks using AI technology for banking support [11]. Here, users can also choose what they would like to be addressed as (their preferred name) and how they would like to be given personalized feedback based on their financial performance.

Because financial CUIs directly target highly sensitive financial information and monetary transactions, we have a domain-specific challenge for developers of chatbots in finance. FinTech comes with its own unique sets of challenges for both innovators and users. These include threats to the security of applications [16] and attacks on services [12]. When individuals' finances are at stake, functionality may take priority over conversational features. While FinTech continues to grow exponentially [14], usable trustworthy CUI research in this area lags behind and current progress is mostly driven by industry. Furthermore, financial decisions have emotional links; money has always been an emotionally-charged subject [2]. Research shows that individuals treat financial data differently from other kinds of personal data, showing greater unwillingness to share the former online [13, 18]. Banks leveraging FinTech recognize that decision-making when it comes to money involves investigation into users' feelings, which does not appear to follow predictions of existing rational information processing models [8]. In light of this, we cannot assume that the same CUI features are viewed as trustworthy across all domains.

In the context of FinBot and FinTech, trust is the willingness of customers to abandon control over the actions performed by the bot and rely on the automation behaving as they expect. Without trust in FinTech, fully acceding to these automated financial processes will be a challenge. Below we report on a preliminary study that has implications in designing trustworthy FinBots. We draw key findings in understanding and identifying the parameters of appropriate CUI design in the financial domain.

## METHOD

*Vignette Emma versus XRO23*. This study compares two chatbot designs, one with conversational and social features named Emma against a more technical bot named XRO23. The study was designed following experimental design guidelines for user-studies [5]. We chose a vignette design to investigate what features linked to FinBots influences trust levels. Technical features were constant in both vignettes, where both chatbots were described as Easy to Use, Secure, Privacy-focused by design, Safe, Speedy, Time-saving, Predictable, Accurate in understanding users' messages (Accuracy), Relevant in Content, Flexible, having a Controllable level of Automation (Control), and offered by a Bank the participants already trust.

For Emma, we included features such as a human name, the ability to respond empathetically and give encouraging statements, active listening skills, the ability to personalize, such as using the user's preferred name and politeness, including being able to turn-take and make friendly small talk. Participants were also explained to that FinBots are pre-programmed to digitally assist users with answers to frequently asked questions and perform actions such as updating users' address and checking their account balance.

*Participants.* $N = 410$ UK participants who were recruited from Prolific Academic took part in this study. The average age was $M = 33.04$ years old ($SD = 11.71$), with $n = 295$ females and $n = 115$ males. $N = 361$ of our participants were Caucasian, $n = 13$ were mixed race, $n = 25$ were of Asian origin, and $n = 8$ identified as Black and the rest were of other ethnic origins.

*Procedure.* Participants were randomly assigned to one of the two conditions ($n = 219$ for Emma and $n = 191$ for XR023), for good design practice. After reading either vignette, they were asked two open-ended free-form questions: Q1 `From the vignette, name all the traits, characteristics and behaviors of the chatbot described in the vignette that helps you trust it?' and Q2 `Apart from the chatbot traits and behaviors mentioned in the vignette, name other elements of a chatbot that would help you trust it more in general?'.

We analyzed these responses through conventional line-by-line coding, with an iterative process of generating a master codebook and coding of all response units. Each unit was independently coded by two independent coders, with high agreement between both coders (Cohen's Kappa = .96, p <.000 for Q1, and Cohen's Kappa = .90, p <.000 for Q2). This approach is similar to individuals' evaluation of privacy attitudes in previous research [6].

**Vignette XRO23:** Imagine a reputable bank that you regularly use have developed a financial text chatbot called XRO23. XRO23 is a financial expert that can assist you. It can analyze your transactions and identifies your regular income, rent, bills and daily spend. Using this and other factors like your available balance, XRO23's algorithm can run every few days and calculates an afford- able amount to set aside for you automatically. XRO23 was built with your security and privacy in mind, safely encrypts data and you are in control (i.e., to set more or less money aside). XRO23 is fast in giving you the information you need, saves your time, and is very accurate in understanding the messages you type to it. The relevancy of its content is high. Chatbot XRO23 is easy to use, predictable, flexible and gives quality results.

**Vignette Emma:** Imagine a reputable bank that you regularly use have developed a financial text chatbot called Emma. Emma is a financial expert who can assist you. She can analyze your transactions and identifies your regular income, rent, bills and daily spend. Using this and other factors like your available balance, Emma's algorithm can run every few days and can calculate an affordable amount to set aside for you automatically. Emma was built with your security and privacy in mind, safely encrypts data and you are in control (i.e., to set more or less money aside). Emma is polite and can be personalized. She can use your preferred name if you want to, and also understands your feelings when you interact with her. She takes her turn to respond appropriately to the conversation, acknowledges and reacts to your feelings accordingly. She also encourages you when you need it by saying things like "You're doing great, carry on", or "Don't be sad, you didn't have better options". Emma is fast in giving you the information you need, saves your time, and is very accurate in understanding the messages you type to her. The relevancy of her content is high. Emma is easy to use, predictable, flexible and gives you quality results.

## RESULTS AND DISCUSSION
### Q1: Privacy and Security are Most Desired
While traditional banks have built a long-standing legacy with handling customer data confidentially and not selling this information to other parties [7], the FinTech sector may have much to prove with regards to the integrity of their data security and information discretion practices. This is reflected in our first key finding, that is, Privacy and Security are the top two most-named features whether participants were presented with the conversational bot Emma or XR023 — in a context where people may still be unaccustomed to the chatbots supporting their finances.

Financial sites and applications are also often uniquely targeted by bad actors (i.e., fraud, malevolent activities and scams) compared to other domains[2]. Our participants may realize this and consider threats to their finances of great importance, thereby requiring Privacy and Security as essential requirements for FinBots over social and conversational features, as compared to bots in other domains. Privacy and Security are followed by Control and Accuracy as depicted in Figure 1.

[2] https://thefintechtimes.com/why-financial-organisations-are-a-prime-target-for-cyber-attacks

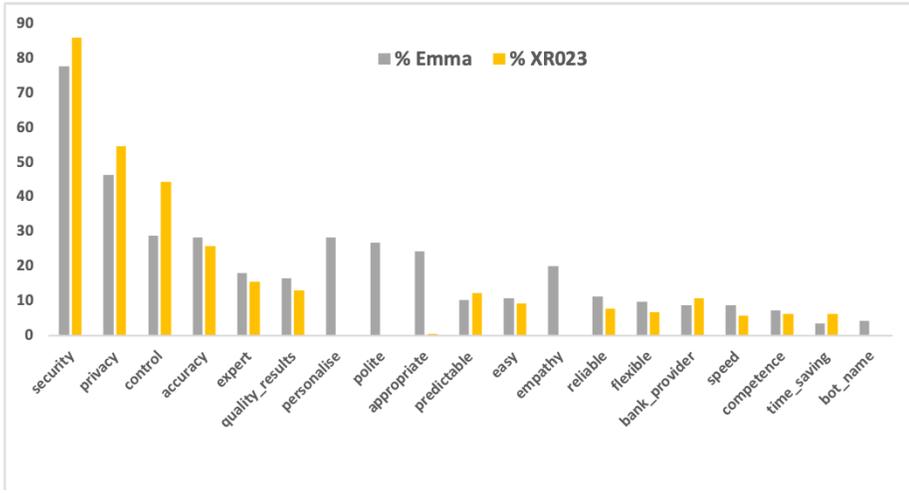

**Figure 1: Vignette features contributing to trust (% participants by condition)**

**Q2: Privacy and Security First Before Conversational Experiences**
Our second key finding is that Privacy and Security are *still* the most desired FinBot features, even after being asked to name features outside of the vignette (see e.g. responses on the next page). This was followed by Social Endorsement, Human-like Conversational Style and Ethical qualities (ethical such as proof of reliability, transparency, or accountability), as shown in Figure 2.

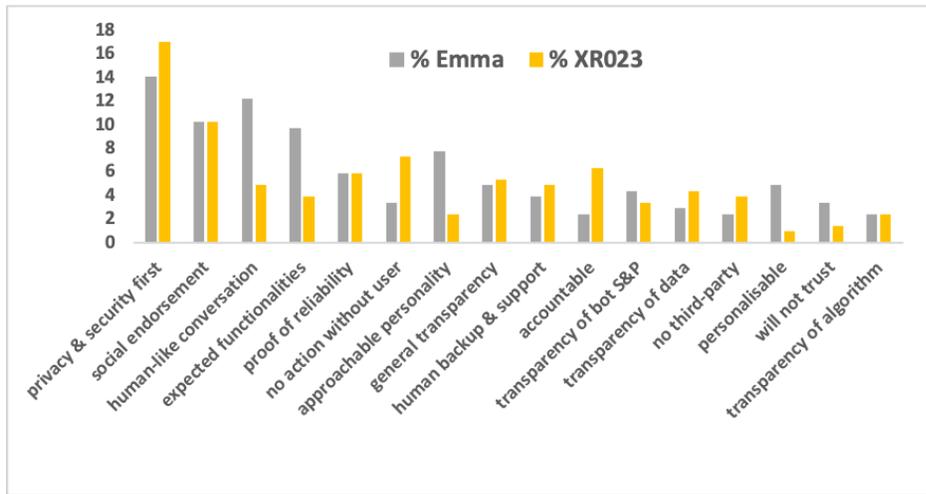

**Figure 2: Additional desired features for trust (% participants by condition**

That said, a high number of participants desired a chatbot to converse like a human after Privacy and Security concerns have been addressed. They wanted a bot that replies in a similar way to a human conversation and objected to automated messages or repetitive stock phrases (e.g., as explained by P259, "For me, I would be more likely to trust a chatbot if I didn't feel like it was a chatbot. By that I mean, if the responses seem pre-programmed or scripted, I am less likely to feel like I have a rapport and therefore less likely to trust"). Participants specified that the bot should exercise tailored responses and be non-generic in its answers. They desired a FinBot that adapts to their speech patterns. Further, participants preferred a bot that is able to understand and adopt colloquialism, informal language and abbreviations (e.g., "Recognizes coloquialisms [sic]", P277).

These findings have practical implications for designers and developers, as individuals are shown to welcome financial technology but at the same time are wary of privacy and security risks [1]. Keeping in mind that we used a vignette study design and since FinBots may access users' financial and private information, users' concerns may increase when actually interacting with the bots. If vulnerabilities pertaining to IT systems and misuse of data can be eliminated or at least mitigated, this can be a key resource in gaining and maintaining trust in FinBots.

Firstly, this means ensuring a privacy- and security-by-design implementation as baseline technology. An example of this would be integrating two-factor authentication methods. Secondly, this translates to security and privacy being demonstrated in financial CUIs and be promoted ethically to users. In order to demonstrate the aspects of system trustworthiness and other ethical FinTech qualities [17], these features should be communicated appropriately or displayed more prominently to provide user assurance. This can take the shape of providing a site verification badge, or being accredited by reliable independent reviews, while protecting against giving users a false sense of security and privacy. In parallel, users can be provided a support service to engage in more secure behaviors as a way to keep them digitally safe (in particular where users appreciate support in engaging with safe technologies [4]). This can take the shape of display icons or tooltips or being provided prompts to regularly change their passwords. Thirdly, to further instill trust in usable FinBots, we propose that conversational features be implemented in financial CUIs (i.e., messages that are clearly not automated or repetitive stock phrases, be provided responses that make grammatical sense, able to understand slang and abbreviations).

Given that banking and finance affect most, if not all, members of society, innovation in this field is likely to have broad and diverse impacts. Such impacts might be positive (i.e., opening up financial services to unbanked or underbanked populations) or negative (i.e., creating new opaque systems through which access to finance is determined). With the growth of FinTech and increased reliance on technology recently, this paper is timely. We provide rich insights to be able to build an attractive application that users will use, that adheres to their privacy and security expectations, while we simultaneously synthesize findings in the FinTech realm.

The research reported in this paper was funded in part by UKRI under grant EP/R033595 FinTrust: Trust Engineering for the Financial Industry.